\documentclass{astlb}

\usepackage[T2A]{fontenc}
\usepackage[utf8]{inputenc}
\usepackage{lscape}
\usepackage{xspace} 
\usepackage{url}
\usepackage{amsmath}
\usepackage{amssymb}

\usepackage[
  hyperfootnotes=false,
  colorlinks=true,
  linkcolor=blue,
  citecolor=blue,
  urlcolor=blue
]{hyperref}
\usepackage{graphicx}
\usepackage{graphicx}
\usepackage{float}
\usepackage[caption=false]{subfig}

\begin{document}

\journalinfo{(2025)}{51}{8}{19}[29]

\title{A Search for Hard X-ray/Soft $\gamma$-ray Emission from  SPT-CL J2012-5649 (Abell 3667) Using \textit{INTEGRAL}/ISGRI}

\author{Siddhant Manna\address{1}\email{ph22resch11006@iith.ac.in},
Shantanu Desai\address{1},
Roman A. Krivonos\address{2} \\
\addresstext{1}{\it Department of Physics, IIT Hyderabad, Kandi, Telangana 502284, India}
\addresstext{2}{\it Space Research Institute (IKI), 84/32 Profsoyuznaya str., Moscow 117997, Russian Federation}
}

\shortauthor{Manna et al.}

\shorttitle{INTEGRAL view of Abell 3667}

\submitted{11.06.2026 \\
Revised June 15, 2026; Accepted June 22, 2026}

\begin{abstract}
We present a search for hard X-ray/soft $\gamma$-ray emission from the merging galaxy cluster SPT-CL J2012-5649 (Abell~3667) using archival INTEGRAL/ISGRI observations. This cluster located at $z=0.0556$ hosts prominent radio relics associated with merger-driven shocks, suggesting the presence of relativistic electrons capable of producing inverse-Compton (IC) emission in the hard X-ray to soft $\gamma$-ray regime. We searched for emission in the 30--300~keV energy range using the INTEGRAL Off-line scientific analysis software with a total effective exposure of 2817~s. No significant emission was detected at the cluster position in the aforementioned energy interval. The extracted ISGRI spectrum is consistent with pure background, and no physically meaningful model parameters can be constrained. From the mosaic variance maps, we derive a $3\sigma$ upper limit of $F_{30-300\,\mathrm{keV}} < 3.63 \times 10^{-10}\,\mathrm{erg\,cm^{-2}\,s^{-1}}$. This limit rules out bright IC scenarios and constrains the efficiency of merger-driven particle acceleration in SPT-CL J2012-5649. Our results provide the most stringent soft $\gamma$-ray constraint on this system to date and highlight the need for next-generation hard X-ray missions, such as HEX-P or eXTP, to probe IC emission at theoretically predicted levels in merging clusters.

\end{abstract}

\section{Introduction}
Galaxy clusters are the most massive gravitationally bound systems in the Universe, comprising hundreds to thousands of galaxies embedded in a hot, diffuse intracluster medium (ICM)~\citep{Murase2013,Condorelli2023}. They serve as powerful laboratories for testing cosmology~\citep{Kravtsov2012,Allen2011,Vikhlininrev} and fundamental physics~\citep{Bohringer16,Desai18,BoraDesaiCDDR,BoraDM,Boraalpha}. The extreme astrophysical environments present in clusters including merger-driven shocks, turbulence, and large-scale particle acceleration are expected to produce non-thermal emission across the electromagnetic spectrum~\citep{Feretti}. Although clusters have been thoroughly observed from radio to  X-ray wavelengths~\citep{Feretti,Wik14}, a definitive detection of gamma-ray emission has remained elusive for decades.

However, in recent years, significant progress has emerged from systematic searches with the Fermi Large Area Telescope (LAT). In our earlier work~\citep{Manna2024}, we conducted a targeted analysis of 300 clusters selected from the $2500~\mathrm{deg}^2$ SPT-SZ survey~\citep{Bleem15,Bocquet19} using 15 years of LAT data. This survey uncovered a single statistically significant gamma-ray excess: a $6.1\sigma$ detection from SPT-CL~J2012$-$5649~\citep{Manna2024}. However, the broad point-spread function of Fermi-LAT complicates interpretation, leaving open the possibility that the emission originates not from diffuse ICM processes but from unresolved point sources such as radio galaxies. 

SPT-CL~J2012$-$5649 is particularly noteworthy because it coincides with the well-studied merging cluster Abell~3667, located at $z = 0.055$ with mass $M_{500} \approx 5\times10^{14}\,M_{\odot}$ and angular radius $\theta_{200} = 0.19^\circ$~\citep{Manna2024}. This system hosts prominent merger-driven shocks and spectacular radio relics observed by ATCA and MeerKAT~\citep{Omiya,Meerkat,deGaasperin22,Carretti13,Riseley}, indicating active cosmic-ray acceleration. The LAT emission is confined within $\sim 0.2\,R_{200}$ and detected only in the 1--10~GeV band~\citep{Manna2024}. Yet, MeerKAT data reveal several compact radio sources within a few arcminutes of the cluster center~\citep{deGaasperin22}, and cross-matching with the SUMSS survey identifies additional potential gamma-ray emitters~\citep{Mauch03}. This diffuse-versus-point-source ambiguity, familiar from the cases of Virgo and Perseus~\citep{Abdo09,Abdo2}, motivates an independent high-energy verification.

To clarify the nature of the emission, we have pursued multi-wavelength and multi-instrument follow-up studies. Archival COMPTEL data show no evidence of MeV emission~\citep{Manna2024b}. A dedicated analysis with the DArk Matter Particle Explorer (DAMPE)~\citep{Chang2017,Chang2014,Ambrosi2019} using point-source, radial-disk, and radial-Gaussian models over 3~GeV--1~TeV likewise revealed no significant signal within $R_{200}$, yielding 95\% C.L. upper limits of $\sim10^{-6}$--$10^{-4}\,\mathrm{MeV\,cm^{-2}\,s^{-1}}$ consistent with LAT measurements~\citep{Manna2024c}. Complementary stacking of 16.4 years of LAT data produced a strong cumulative signal from SPT-SZ clusters with individual TS$<9$ (combined TS $=75.2$, $\sim 8.4\sigma$), suggesting a population-level gamma-ray component~\citep{Manna2024d}.

Despite these developments, the hard X-ray to soft gamma-ray regime (20~keV--10~MeV) remains largely unexplored for galaxy clusters. This energy range is crucial for distinguishing hadronic from leptonic emission and for testing inverse-Compton models of cosmic-ray populations. The INTEGRAL mission provides unique sensitivity in this band, combining superior angular resolution compared to Fermi-LAT with coverage extending from $\sim 20$~keV into the MeV domain. In this work, we present the first INTEGRAL/ISGRI analysis of SPT-CL~J2012$-$5649 (Abell~3667). Our goals are (1) to test whether the LAT-detected excess can be independently confirmed with an instrument of higher angular resolution, and (2) to constrain non-thermal emission in the poorly studied 20~keV--300~keV range, thereby placing new limits on particle acceleration in merging clusters. 

While the sensitivity of INTEGRAL/ISGRI to galaxy clusters has been explored through simulations~\citep{Goldoni}, observational searches have been limited. \citet{Krivonos2022} analyzed the Ophiuchus cluster using INTEGRAL/ISGRI data spanning 2003--2009 and reported marginal evidence ($5.5\sigma$) for a possible non-thermal component above 60~keV. However, they concluded that systematic uncertainties in decomposing the thermal and non-thermal emission at energies above 20~keV precluded a definitive detection of the non-thermal excess. Similarly,~\citet{Lutovinov2008} conducted a multi-instrument study of the Coma Cluster combining INTEGRAL, RXTE, and ROSAT data over a broad energy range (0.5--107~keV). They found that the Coma spectrum could be well described by thermal plasma emission with $T = 8.2$~keV, with only marginal detection ($\sim1.6\sigma$) of the cluster in the hard 44--107~keV band. From their analysis, they derived a 20--80~keV flux for a possible non-thermal component of $(6.0 \pm 8.8) \times 10^{-12}$~erg~cm$^{-2}$~s$^{-1}$, consistent with no detection. 

For Abell~3667 specifically, previous hard X-ray measurements have reported flux levels at $\sim 10^{-12}$--$10^{-11}$~erg~cm$^{-2}$~s$^{-1}$. Using combined \textit{XMM-Newton} and \textit{Swift}/BAT data, it was found the 50--100~keV emission could be described either by a hot thermal component or by a power-law model with photon index $\Gamma \approx 1.8$, corresponding to a flux of $3.0^{+4.2}_{-0.7} \times 10^{-12}$~erg~cm$^{-2}$~s$^{-1}$~\citep{Ajello2010}. More recently, deep \textit{NuSTAR} observations of the central region~\citep{Mirakhor2025} reported a 20--80~keV flux of $9.5^{+0.5}_{-5.5} \times 10^{-12}$~erg~cm$^{-2}$~s$^{-1}$ when fitted with a power-law component of photon index $\Gamma = 1.6^{+0.1}_{-0.2}$, although their analysis favored a multi-temperature thermal interpretation. The brightest \textit{Chandra}-resolved point source in the field contributes only $2.2^{+1.5}_{-1.1} \times 10^{-13}$~erg~cm$^{-2}$~s$^{-1}$ in the 20--80~keV band, more than an order of magnitude below the reported excess, indicating that the hard X-ray emission is unlikely to be dominated by a single unresolved AGN.
\textit{Suzaku} observations (0.5--40~keV) also investigated the presence of non-thermal hard X-ray emission in Abell~3667 and found that the $>10$~keV emission near the cluster center is better explained by a very hot ($kT > 13$~keV) thermal component rather than inverse-Compton emission. In the northwestern radio relic, no significant non-thermal signal was detected, with a 90\% confidence level upper limit of $7.3 \times 10^{-13}$~erg~cm$^{-2}$~s$^{-1}$ in the energy range of 10--40~keV~\citep{Nakazawa}.

More recently,~\citet{Wik2014} presented deep NuSTAR observations of the Bullet cluster, one of the most promising candidates for inverse-Compton detection due to its extreme merger state and prominent shock features. Despite NuSTAR's unprecedented hard X-ray focusing capability and 266~ks exposure, they found no convincing evidence for non-thermal emission in the 3--30~keV band. This result, from a focusing telescope with superior background characterization compared to coded-mask instruments, underscores the challenge of detecting inverse-Compton emission even in the most favorable systems. These studies collectively highlight the difficulty of detecting non-thermal hard X-ray emission from galaxy clusters, even for massive, nearby systems with strong merger signatures. To our knowledge, the present work represents the first dedicated INTEGRAL analysis of a radio relic hosting merging cluster specifically selected for its strong indicators of ongoing particle acceleration. Unlike previous serendipitous or archival studies, Abell~3667 was targeted precisely because its prominent radio relics~\citep{deGaasperin22} and reported GeV-band excess~\citep{Manna2024} make it one of the most promising candidates for detectable inverse-Compton emission from shock-accelerated electrons. This study represents an essential step toward establishing whether galaxy clusters are genuine gamma-ray sources and toward understanding the physical origin of their high-energy emission.

The remainder of the paper is organized as follows. Section~\ref{sec:methods} describes the \textit{INTEGRAL} observations, data reduction, and analysis procedures. Section~\ref{sec:results} presents the imaging and spectral results. We summarize our conclusions in Section~\ref{sec:discussion}. For the rest of the manuscript we shall use Abell 3667 to refer to SPT-CL J2012-5649.

\section{Data Analysis}
\label{sec:methods}

The search for high energy emission from Abell~3667 straddling the boundary between hard X-rays and soft $\gamma$-rays was conducted using the ISGRI detector of the IBIS instrument~\citep{Ubertini2003} onboard \textit{INTEGRAL}~\citep{Winkler2003}. ISGRI is sensitive in the 15~keV--1~MeV energy range, with an angular resolution of $\sim12'$ (FWHM) and a field of view of $29^\circ \times 29^\circ$~\citep{Lebrun2003}, making it well-suited for detecting faint emission in the soft $\gamma$-ray regime. All data were processed using the \textit{INTEGRAL} Off-line Scientific Analysis software package (OSA v11.2)~\citep{Courvoisier2003}, following standard procedures outlined in the OSA User Manual\footnote{\url{https://www.astro.unige.ch/integral/analysis}}. The coded-mask imaging design of INTEGRAL/ISGRI provides simultaneous reconstruction of multiple sources while ensuring stable background modeling, enabling robust searches for weak high-energy signals. 

All publicly available Science Windows (ScWs) covering the position of Abell~3667 (RA~=~$303.14^\circ$, Dec~=~$-56.84^\circ$, J2000) were retrieved from the INTEGRAL Science Data Archive\footnote{\url{https://www.isdc.unige.ch/integral/archive}}.  After applying standard quality filters to exclude time intervals affected by elevated particle background, a total of 16 Science Windows were found within a $5^\circ$ radius of the cluster, as shown in Table~\ref{tab:scw_all}. For the primary analysis, we restrict the dataset to the 
four Science Windows with the smallest angular separations 
from the cluster center ($\sim$2.5$^\circ$--2.8$^\circ$, 
or equivalently 150--170~arcmin), for which Abell~3667 
lies well within the fully coded field of view of ISGRI. 
At larger off-axis angles ($\gtrsim$3$^\circ$), the ISGRI 
coding fraction decreases rapidly and background 
systematics become increasingly dominant, substantially 
reducing sensitivity to faint and diffuse emission. 
Including the remaining 12 Science Windows at such 
large off-axis angles would therefore not improve, and 
could potentially degrade, the robustness of the derived 
flux constraints. The four selected Science Windows 
provide an effective ISGRI exposure of 2817~s at the 
cluster position. We note that this limited exposure 
reflects the absence of dedicated pointed INTEGRAL 
observations of Abell~3667 in the public archive; all 
available Science Windows are serendipitous observations 
in which the cluster falls within the ISGRI field of 
view rather than at the primary pointing position.

As a consistency check, we repeated the full ISGRI 
analysis including all 16 publicly available Science 
Windows within a $5^\circ$ radius of Abell~3667. For 
this extended dataset, we restricted the energy range 
to 30--100~keV, where the ISGRI effective area is at 
its highest and the instrumental response is best 
controlled, thereby avoiding the rapidly decreasing 
sensitivity and increasing background systematics above 
$\sim$100~keV. The results of this extended analysis, 
which are fully consistent with the primary four-ScW 
results, are presented and discussed in 
Appendix~\ref{app:catalog_check}.

A unified observation group was constructed using the \texttt{og\_create} task, which associates the selected ScWs with the appropriate instrument characteristics, auxiliary files, and reference catalog entries. The subsequent imaging analysis was configured to produce sky mosaics in four logarithmically spaced energy bands: 30--53, 53--95.5, 95.5--168, and 168--299.5~keV. These bands were chosen to optimize sensitivity given the energy dependent effective area of ISGRI while maintaining adequate photon statistics in each band. The chosen energy range reflects both instrumental limitations and scientific considerations. The low energy response of ISGRI has degraded during the mission due to an increase in the number of inactive pixels and evolving threshold settings. As documented in the OSA User Manual~\citep{Courvoisier2003}, the detector’s efficiency becomes negligible below $\sim25$~keV after approximately Revolution~1600 (mid-2015). The ISDC therefore recommends avoiding energies below $E_{\rm min}+5$~keV, implying a practical lower limit near 30~keV for data in our observing epochs. The upper bound of 300~keV was selected because the effective area and sensitivity decline steeply above this energy, providing little additional constraining power for faint sources such as galaxy clusters. For each energy interval, intensity, variance, significance, and exposure maps were produced.

The image reconstruction process began at the energy-corrected level (\texttt{startLevel = COR}) and proceeded through the image reconstruction level (\texttt{endLevel = IMA2}). The default \textit{INTEGRAL} reference catalog (\texttt{CAT\_refCat = \$ISDC\_REF\_CAT[ISGRI\_FLAG>0]}) was adopted, which contains all sources detected by ISGRI in public data prior to the catalog release. Only ISGRI data were processed (\texttt{SWITCH\_disablePICsIT = yes}), and no additional user-defined good time intervals were applied beyond the quality filters described above. The standard bad time interval exclusion criteria (\texttt{SCW1\_GTI\_BTI\_Names}) were implemented to reject periods affected by instrumental or environmental anomalies, as detailed in the OSA documentation.

Sky images were generated in each of the four energy bands using the \texttt{ii\_skyimage} task. Source detection was performed using a combined approach (\texttt{OBS1\_SearchMode = 3}) that searched for both catalog sources and up to 50 additional bright sources in the field of view (\texttt{OBS1\_ToSearch = 50}), with a minimum detection significance threshold of $5\sigma$ for new sources (\texttt{OBS1\_MinNewSouSnr = 5}). The positions of all catalog sources were fitted during the analysis (\texttt{OBS1\_SouFit = 0}), except for sources with \texttt{ISGRI\_FLAG = 2}, 
whose positions are known to better than $3''$ 
accuracy and were therefore held fixed at their 
catalog values. Background subtraction was performed using the standard ISGRI background maps provided by the IBIS team, with normalization calculated after excluding pixels affected by bright catalog sources (\texttt{brSrcDOL = "\$ISDC\_REF\_CAT[ISGRI\_FLAG2==5 \&\& ISGR\_FLUX\_1>100]"}).

After processing individual ScW images, a mosaic image was constructed by combining all four ScWs (\texttt{OBS1\_DoPart2 = 1}). The mosaic was generated with pixel spreading enabled (\texttt{OBS1\_PixSpread = 1}), which distributes source photons around a single central peak to enable sub-pixel positioning and thereby improve source localization accuracy. To verify the robustness of our results against the choice of energy binning, we repeated the entire analysis using both 2 and 8 logarithmically spaced energy bins. The detection results remained consistent across all binning schemes, confirming that our fiducial choice of four energy bands provides an optimal balance between energy resolution and photon statistics. In addition, motivated by the rapid decline of the ISGRI effective area above $\sim$100~keV, we performed a dedicated analysis in a reduced 30--100~keV energy band; this also yielded no statistically significant emission, with all signals remaining below the $2\sigma$ level.

Following image reconstruction, spectral extraction 
was performed for the cluster position using the 
\texttt{spe\_pick} tool, processing data from the 
binning level (\texttt{startLevel = BIN\_S}) through 
the spectral extraction level (\texttt{endLevel = SPE}), 
with the same background maps as the imaging analysis 
to ensure consistency. The extracted spectra confirmed 
the absence of statistically significant emission at 
the cluster position, consistent with the image-based 
upper limits derived in Section~\ref{sec:upperlimits}.

The significance of potential detections was evaluated using the Test Statistic (TS), which quantifies the likelihood ratio between models with and without a source at the target position~\citep{Manna2024}. According to Wilks' theorem~\citep{Wilks1938}, the TS approximately follows a $\chi^2/2$ distribution with degrees of freedom equal to the number of free parameters in the source model~\citep{Mattox1996}. For ISGRI imaging, the detection significance in standard deviations is given by $\sigma = \sqrt{\mathrm{TS}}$, where the TS values are computed directly from the significance maps produced by the \texttt{ii\_skyimage} task. We adopted a $5\sigma$ threshold (corresponding to TS~$\approx$~25)~\citep{Lyons2013} for claiming source detection in the imaging analysis. As discussed later, no statistically significant emission from Abell~3667 was detected in any of the four energy bands. The maximum excess significance occurs in the 30--53~keV band, with a value of $1.96\sigma$, which is well below the threshold required for a credible detection. Consequently, flux upper limits were derived using a purely image-based approach, as described in Section~\ref{sec:upperlimits}.

\section{Results}
\label{sec:results}
We present the results of our INTEGRAL/ISGRI analysis of Abell~3667, a dynamically active merging cluster at redshift $z = 0.0556$ that hosts prominent radio relics associated with merger-driven shocks and ongoing particle acceleration~\citep{deGaasperin22,Riseley}. Our objective is to search for non-thermal soft $\gamma$-ray emission in the 30--300~keV band, which would provide direct evidence for inverse-Compton (IC) radiation from relativistic electrons in the intracluster medium (ICM). Such emission is predicted by leptonic and hadronic models of cosmic-ray interactions in clusters and would complement the tentative GeV-band excess previously reported with Fermi-LAT~\citep{Manna2024}.

The data reduction and imaging analysis were performed using OSA~11.2, following the procedures described in Section~\ref{sec:methods}. 
A systematic search of the ISDC public archive identified four Science Windows (ScWs) covering the position of Abell~3667 (RA $= 303.14^\circ$, Dec $= -56.84^\circ$, J2000) over the mission lifetime between 2003 October~16 and 2025 November~2. The properties of these observations are summarized in Table~\ref{tab:scw_selection}. With angular offsets of 150--170~arcmin, the cluster lies well within the fully coded field of view of ISGRI for all selected pointings. 
The total effective exposure accumulated from these observations is 2817~s, distributed across two widely separated epochs.

\begin{table*}[ht]
\centering
\caption{Science Windows used in the INTEGRAL/ISGRI analysis of Abell~3667. Angular separations are computed between the ScW pointing center and the cluster position (RA~=~$303.14^\circ$, Dec~=~$-56.84^\circ$, J2000).}
\label{tab:scw_selection}
\begin{tabular}{lccc}
\hline\hline
\textbf{ScW ID} & \textbf{Pointing Center (RA, Dec)} & \textbf{good\_isgri (s)} & \textbf{Separation (arcmin)} \\
\hline
091100160010  & $(302.886^\circ,\,-54.290^\circ)$ & 506 & 153.3 \\
091100170010 & $(302.868^\circ,\,-54.246^\circ)$ & 688 & 156.0 \\
186500370010  & $(307.235^\circ,\,-58.605^\circ)$ & 779 & 168.5 \\
186500380010  & $(306.982^\circ,\,-58.417^\circ)$ & 844 & 155.4 \\
\hline
\hline
\end{tabular}
\end{table*}

\begin{table}
\centering
\caption{Known ISGRI sources within the field of view of Abell~3667 observations, sorted by angular separation. Separations are computed from the cluster center at RA~=~$303.14^\circ$, Dec~=~$-56.84^\circ$ (J2000). Only SWIFT~J2012.0-5648 lies within the ISGRI PSF ($\sim 12'$).}
\label{tab:isgri_sources}
\begin{tabular}{lccc}
\hline\hline
Source Name & RA & Dec & Sep. \\
            & (deg) & (deg) & (arcmin) \\
\hline
SWIFT J2012.0-5648        &  303.06 &  -56.82 &    2.8 \\
SWIFT J2018.4-5539        &  304.43 &  -55.68 &   81.9 \\
ESO 141-55                &  290.31 &  -58.67 &  424.4 \\
IGR J19387-6502           &  294.74 &  -65.04 &  548.6 \\
IGR J19386-4653           &  294.65 &  -46.89 &  673.8 \\
ESO 140-43                &  281.22 &  -62.36 &  738.4 \\
SWIFT J1839.1-5717        &  279.76 &  -57.25 &  759.6 \\
IGR J21024-4608           &  315.63 &  -46.11 &  792.5 \\
ESO 103-35                &  279.58 &  -65.43 &  846.5 \\
IGR J18244-5622           &  276.08 &  -56.37 &  888.2 \\
IGR J20526-4320           &  313.16 &  -43.35 &  894.7 \\
\hline
\end{tabular}
\end{table}

\begin{figure*}
\centering
\includegraphics[width=0.75\textwidth]{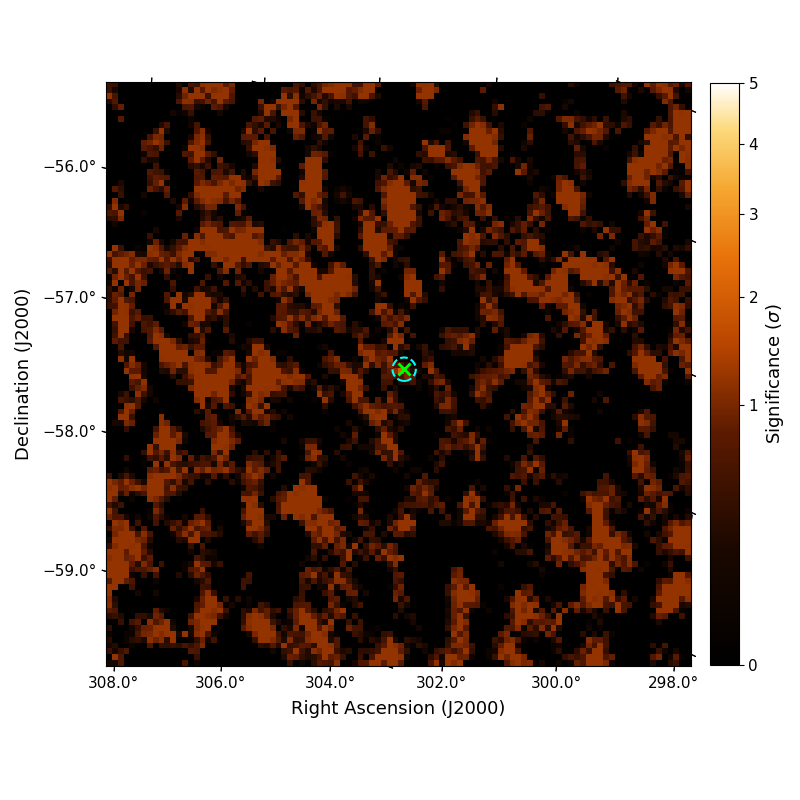}
\caption{ISGRI significance map of the Abell~3667 field in Band~1 (30--53~keV). 
The green cross marks the position of Abell~3667 (RA~$= 303.14^\circ$, Dec~$= -56.84^\circ$), 
where a significance of $1.96\sigma$ is detected, consistent with statistical background 
fluctuations and well below the standard $5\sigma$ detection threshold adopted for ISGRI 
source identification. The cyan dashed circle, centred on the source position, represents 
the angular resolution of IBIS/ISGRI ($\sim$12$'$\ FWHM). The color scale represents detection significance in units of standard deviations, displayed with square-root scaling over the range 0--5$\sigma$.}
\label{fig:significance_map}
\end{figure*}

We generated mosaic images in four logarithmically spaced energy bands spanning 30--300~keV and examined the corresponding significance maps, where the significance is defined as the square root of TS, as described in Section~\ref{sec:methods}. The extracted significance values at the cluster center range from $0.1$--$1.96\sigma$ across the four bands, consistent with statistical background fluctuations and well below the standard $5\sigma$ detection threshold adopted for ISGRI source identification (Table~\ref{tab:significance_values}). Figure~\ref{fig:significance_map} shows the Band~1 (30--53~keV) mosaic, which yields the highest significance of $1.96\sigma$ at the cluster position (RA~$= 303.14^\circ$, Dec~$= -56.84^\circ$), marked by a green cross; the remaining bands show lower significance and are not shown as they provide no additional information beyond confirming the non-detection. We therefore place an upper limit on the hard X-ray flux from Abell~3667 in the ISGRI energy range.

To evaluate potential contamination from nearby hard X-ray 
emitters, we cross-matched the field with the IBIS/ISGRI 
1000-orbit source catalog~\citep{Bird2016}. Within a radius 
of $15^\circ$ from the cluster center, 11 cataloged sources 
are present (Table~\ref{tab:isgri_sources}). The closest of 
these is SWIFT~J2012.0$-$5648, located $2.8'$ from the 
center of Abell~3667 and therefore well within the 
$\sim12'$ ISGRI point-spread function.

We note that SWIFT~J2012.0$-$5648, despite its 
proximity to Abell~3667 ($2.8'$, within the ISGRI PSF), 
does not pose a contamination concern for our analysis. 
The source is cataloged as a persistent emitter only in 
the 17--30~keV band at $\sim$1.2~mCrab~\citep{Bird2016}, 
with no significant emission reported above 30~keV. Since 
our entire analysis is conducted in the 30--300~keV band, 
the source falls outside our energy range regardless of 
its flux level or the exposure time of our observations. 
In our 30--300~keV mosaics, no emission is detected at 
the position of SWIFT~J2012.0$-$5648 (significance 
$< 2\sigma$ in all energy bands). Consequently, any 
potential contribution from this source to the cluster 
flux upper limits derived in this work is negligible.

The next nearest source, SWIFT~J2018.4$-$5539, is offset 
by $82'$, well outside the ISGRI point-spread function. 
All remaining cataloged sources are located at angular 
separations exceeding $7^\circ$. We therefore conclude 
that contamination from known ISGRI sources does not 
affect our search for emission from Abell~3667.

\begin{table*}
\centering
\caption{Significance values at source positions in the ISGRI significance maps. 
Values are extracted at the positions of Abell~3667 and the two nearest catalog sources.}
\label{tab:significance_values}
\begin{tabular}{lcccc}
\hline\hline
Source & Band 1 & Band 2 & Band 3 & Band 4 \\
       & (30--53 keV) & (53--95.5 keV) & (95.5--168 keV) & (168--299.5 keV) \\
\hline
Abell 3667                     &   1.96 &   1.58 &  0.1 &   0.45 \\
SWIFT J2012.0-5648             &   0.85 &   1.21 &  0.2 &   0.48 \\
SWIFT J2018.4-5539             &   0.51 &   0.59 &  0.1 &   0.64 \\
\hline
\end{tabular}
\end{table*}

Spectral extraction at the cluster position confirmed 
the absence of statistically significant emission in 
all four energy bands, consistent with the 
image-based results presented above.

\subsection{Flux Upper Limits}
\label{sec:upperlimits}

Given the absence of a statistically significant detection, we derive flux upper limits using an image-based approach. No significant excess emission is detected from Abell~3667 in the 30--300~keV energy band. The statistical uncertainty at the cluster position was estimated directly from the mosaic variance image. The $1\sigma$ count-rate uncertainty was obtained as
\[
\sigma_{\rm cnt} = \sqrt{V},
\]
where $V$ is the variance value at the pixel corresponding to the 
cluster coordinates. This yields
\[
\sigma_{\rm cnt} = 0.622~\mathrm{counts~s^{-1}},
\]
corresponding to a conservative $3\sigma$ upper-limit count rate of
\[
C_{3\sigma} = 1.866~\mathrm{counts~s^{-1}}.
\]

To convert the count-rate upper limit into a physical energy flux, we employed a Crab Nebula-based calibration derived directly from the mosaic intensity image in the same energy band. The Crab Nebula was processed using the identical analysis pipeline and energy selection as Abell~3667, and its count rate was measured from the mosaic image. Assuming a Crab spectrum of the form
\[
N(E) = 10\,E^{-2.1}~\mathrm{ph~cm^{-2}~s^{-1}~keV^{-1}}
\]
\citep{Krivonos2010}. The flux of any source is then estimated as
\[
F_{\rm source} = \frac{C_{\rm source}}{C_{\rm Crab}} \times F_{\rm Crab},
\]
where $C_{\rm source}$ and $C_{\rm Crab}$ are the source and Crab 
count rates measured from the same mosaic image under identical 
conditions.

Applying this calibration to the variance-based count-rate upper limit of Abell~3667, we obtain a $3\sigma$ flux upper limit of
\[
F_{3\sigma}(30\!-\!300~\mathrm{keV}) = 3.63 \times 
10^{-10}~\mathrm{erg~cm^{-2}~s^{-1}}~(\approx 64~\mathrm{mCrab}),
\]
which is well within the physically meaningful sensitivity range expected from the image-based method \citep{Krivonos2010}.

This provides an independent constraint on the broadband soft $\gamma$-ray emission from Abell~3667 in the 30--300~keV band using INTEGRAL/ISGRI, complementing existing measurements from \textit{Swift}/BAT, \textit{NuSTAR}, and \textit{Suzaku}. As an independent verification, we compared our results with significance maps generated using an alternative analysis pipeline, based on improved reconstruction methods developed for the INTEGRAL hard X-ray survey~\citep{Krivonos2010b}. 
The survey data and processed sky images  through this pipeline are publicly accessible through the INTEGRAL survey archive~\footnote{\url{https://integral.cosmos.ru/}}.
These independent maps, covering comparable energy bands over the 30--300~keV range, confirm the absence of significant emission at the cluster position, with significance values consistent with our OSA-based analysis within statistical uncertainties.  

Although our derived $3\sigma$ upper limit is significantly higher than the hard X-ray fluxes previously reported from \textit{Swift}/BAT and \textit{NuSTAR}, and is therefore not strongly constraining for the non-thermal emission from Abell~3667, it nevertheless provides an independent constraint using the wide-field INTEGRAL/ISGRI data set. In particular, our analysis extends to higher energies (30--300~keV) compared to the 20--80~keV 
(\textit{NuSTAR}) and 50--100~keV 
(\textit{Swift}/BAT) ranges, making a direct quantitative comparison non-trivial. Such searches remain valuable, since without systematically probing hard X-ray observations it would not be possible to assess whether non-thermal emission is detectable at higher energies or over larger spatial scales.

\subsection{Flux Upper Limits at the Radio Relic Positions}
\label{sec:relic_limits}

Abell~3667 hosts two radio relics located $\sim29'$ (1.8~Mpc) on either side of the cluster center along the merger axis \citep{Finoguenov2010}. The northwest (NW) relic spans $0.6^{\circ}$ (2.3~Mpc) while the southeast (SE) relic spans $0.42^{\circ}$ (1.6~Mpc) \citep{deGaasperin22}.

Following the procedure of Section~\ref{sec:upperlimits}, we extracted $\sigma_{\rm cnt}=0.625$ and $0.6247$~counts~s$^{-1}$ from the 30--300~keV variance mosaic at the NW and SE relic positions, corresponding to
\[
C_{3\sigma,\rm NW}=1.875~\mathrm{counts~s^{-1}}, \qquad
C_{3\sigma,\rm SE}=1.874~\mathrm{counts~s^{-1}}.
\]
Applying the same Crab calibration as Section~\ref{sec:upperlimits}, we obtain
\[
F_{3\sigma,\rm NW} < 3.65\times10^{-10}~\mathrm{erg~cm^{-2}~s^{-1}}\;
\]
\[
F_{3\sigma,\rm SE} < 3.64\times10^{-10}~\mathrm{erg~cm^{-2}~s^{-1}}\;
\]
Both limits are essentially identical to the cluster-center value ($3.63\times10^{-10}$~erg~cm$^{-2}$~s$^{-1}$) consistent with ISGRI's $\sim12'$ angular resolution \citep{Lebrun2003} being comparable to the relic--center separation.

\subsection{Physical Expectations from Radio Data}
\label{sec:ic_prediction}

To place our flux upper limits in physical context, we estimate the expected inverse-Compton (IC) hard X-ray emission from the northwestern radio relic of Abell~3667, following the standard 
framework in which the same relativistic electron population responsible for synchrotron radio emission also upscatters cosmic microwave background (CMB) photons to X-ray energies. We focus on the northwestern relic as it dominates the expected IC emission, being approximately one order of 
magnitude brighter in radio than the southern relic~\citep{deGaasperin22}; the latter is expected to contribute $\lesssim$10\% of the total IC flux.

The ratio of IC to synchrotron flux is given by the ratio of CMB to magnetic field energy densities. Given a relativistic electron population, the IC emissivity is directly proportional to the CMB energy density $U_{\rm CMB}$, while the synchrotron emissivity is proportional to the 
magnetic field energy density $U_B = B^2/8\pi$, leading to \citep{Blumenthal1970, Carilli2002}
\begin{equation}
\frac{F_{\rm IC}}{F_{\rm sync}} = \frac{U_{\rm CMB}}{U_B},
\end{equation}
where in most astrophysical circumstances the background 
photon field is dominated by the CMB \citep{Carilli2002}, 
and at the cluster redshift $z = 0.055$ the CMB energy 
density is \citep{Petrosian2008,Battistelli2002}:
\begin{equation}
U_{\rm CMB} = aT_0^4(1+z)^4 
= 4.19\times10^{-13}(1+z)^4
\approx 5.2\times10^{-13}~\mathrm{erg~cm^{-3}},
\end{equation}
where $a = 7.566\times10^{-15}$~erg~cm$^{-3}$~K$^{-4}$ is the radiation constant and $T_0 = 2.725$~K is the present-day CMB temperature \citep{Fixsen2009}.

For the radio input, we adopt the MeerKAT flux density of the northwestern relic from \citet{deGaasperin22}, who report $S_{1.28\,\rm GHz} = 2.34 \pm 0.12$~Jy (point-source corrected), 
with a spectral index $\alpha = 1.4$ (where $S_\nu \propto \nu^{-\alpha}$). Integrating the synchrotron spectrum over the radio band (10~MHz--10~GHz), we obtain a total synchrotron energy flux of $F_{\rm sync} \approx 5.1\times10^{-12}$~erg~cm$^{-2}$~s$^{-1}$. For magnetic field strengths typical of cluster radio relics, $B \sim 1$--$3~\mu$G, the magnetic energy density is 
$U_B = (0.40\text{--}3.58)\times10^{-13}$~erg~cm$^{-3}$, yielding a CMB-to-magnetic energy density ratio of $U_{\rm CMB}/U_B = 1.45\text{--}13.1$. The predicted IC flux in the 30--100~keV band is therefore:
\begin{equation}
F_{\rm IC}(30\text{--}300~\mathrm{keV}) 
\sim (0.7\text{--}6.6)\times10^{-11}~\mathrm{erg~cm^{-2}~s^{-1}},
\end{equation}
consistent with the hard X-ray excesses reported from \textit{Swift}/BAT and \textit{NuSTAR} observations \citep{Ajello2010, Mirakhor2025}. Our INTEGRAL/ISGRI $3\sigma$ upper limit of 
$3.63\times10^{-10}$~erg~cm$^{-2}$~s$^{-1}$ in the 30--300~keV band is well above the theoretical expectation, confirming that the present dataset lacks the sensitivity to detect or constrain IC emission from Abell~3667.

The most directly comparable INTEGRAL study is that of the Coma cluster by \citet{Lutovinov2008}, who combined INTEGRAL (17--107~keV), RXTE (3--20~keV), and ROSAT (0.5--2~keV) data over the full 0.5--107~keV band. Even for this significantly brighter ($z = 0.023$) and closer system, the broadband spectrum was well described by purely thermal plasma emission at $kT = 8.2 \pm 0.2$~keV. 
In the 44--107~keV band, flux of $(1.8\pm1.1)\times10^{-11}$~erg~cm$^{-2}$~s$^{-1}$ was measured at only $1.6\sigma$ significance, with a 95\% upper limit of $3.3\times10^{-11}$~erg~cm$^{-2}$~s$^{-1}$. Fitting a non-thermal power-law component with fixed 
photon index $\Gamma = 2$ over the broader 20--80~keV band yields a flux of $(6.0\pm8.8)\times10^{-12}$~erg~cm$^{-2}$~s$^{-1}$, consistent with zero at the $0.7\sigma$ level. Point source contamination was excluded down to $7.6\times10^{-12}$~erg~cm$^{-2}$~s$^{-1}$ in the 
44--107~keV band. Even for the Coma cluster INTEGRAL could not conclusively detect non-thermal IC emission above the thermal component. Our non-detection of the more distant Abell~3667 is 
therefore fully expected and physically consistent. A meaningful IC 
constraint for Abell~3667 would require a substantially deeper INTEGRAL exposure.

\section{Discussion and Conclusions}
\label{sec:discussion}
We have conducted a targeted search for hard X-ray/soft $\gamma$-ray emission from the merging galaxy cluster Abell~3667 using archival INTEGRAL/ISGRI observations. Despite a modest effective exposure of 2817~s, the resulting mosaics and spectra are of sufficient quality to place meaningful constraints on non-thermal emission associated with merger-driven particle acceleration. No significant signal is detected at the cluster position in any of the four energy bands between 30 and 300~keV. Moreover, we find no evidence for contamination from nearby high-energy sources, including the cataloged source SWIFT~J2012.0$-$5648 located only $2.8'$ from the cluster center.

The absence of a detectable ISGRI signal from Abell~3667 is consistent with the broader observational picture in which strong non-thermal soft $\gamma$-ray emission remains elusive in galaxy clusters. Using an image-based analysis and instrument-motivated energy bands, we derive $3\sigma$ upper limits of $F_{30\!-\!300\,\mathrm{keV}} < 3.63 \times 10^{-10}$~erg~cm$^{-2}$~s$^{-1}$ ($\sim$64~mCrab). This constraint add Abell~3667 to the growing sample of merging clusters in which no bright inverse-Compton component is detected, despite compelling evidence for merger-driven shocks and efficient particle acceleration inferred from radio observations~\citep{deGaasperin22,Riseley}.

We note that this upper limit is approximately two orders of magnitude higher than the hard X-ray flux levels previously reported from \textit{NuSTAR} and \textit{Swift}/BAT observations ($\sim 10^{-12}$--$10^{-11}$~erg~cm$^{-2}$~s$^{-1}$). Therefore, while our result provides the first constraint in the 30--300~keV band obtained from INTEGRAL/ISGRI using a purely image-based approach, it does not place stringent limits on the lower flux levels inferred from focusing hard X-ray instruments. Instead, our measurement confirms the absence of any bright ($\gtrsim 10^{-10}$~erg~cm$^{-2}$~s$^{-1}$) soft $\gamma$-ray emission from the cluster. Previous \textit{NuSTAR} analyses indicate that the reported hard X-ray excess is more naturally described by multi-temperature thermal emission associated with merger-driven heating; however, a non-thermal inverse-Compton contribution cannot be completely ruled out, particularly given the presence of prominent radio relics and the dynamically complex, turbulent environment of Abell~3667. Our results are therefore consistent with a scenario in which any non-thermal hard X-ray component, if present, lies below the current sensitivity of INTEGRAL/ISGRI.

This result has several implications for interpreting the GeV-band excess previously reported from SPT-CL~J2012$-$5649~\citep{Manna2024}, which is spatially coincident with Abell~3667. If the GeV emission originates from inverse-Compton scattering by relativistic electrons in the intracluster medium, a corresponding hard X-ray/soft $\gamma$-ray component should arise from the same electron population at lower energies. The ISGRI upper limit rules out any such IC component brighter than $\sim 3.63 \times 10^{-10}$~erg~cm$^{-2}$~s$^{-1}$ in the 30--300~keV band. The emission may arise from unresolved point sources particularly radio galaxies, several of which are known to lie in projection near the cluster center~\citep{Manna2024} or from hadronic interactions between cosmic-ray protons and the thermal ICM gas. 

Future observations with next-generation hard X-ray missions will be crucial for resolving the nature of high-energy emission from merging clusters. Instruments such as HEX-P~\citep{Madsen2024} and eXTP~\citep{Zhang2025}, with projected sensitivities an order of magnitude better than INTEGRAL/ISGRI in the 20--200~keV band, will be capable of detecting IC flux levels well below the limits presented here. 

\section{Acknowledgements}
SM gratefully acknowledges the Ministry of Education (MoE), Government of India, for their consistent support through the research fellowship, which has been instrumental in facilitating the successful completion of this work. This research is based on observations with INTEGRAL, an ESA project with instruments and science data centre funded by ESA member states (especially the PI countries: Denmark, France, Germany, Italy, Switzerland, Spain), Czech Republic and Poland, and with the participation of Russia and the USA. We acknowledge the use of public data from the INTEGRAL Science Data Centre (ISDC).

\bibliographystyle{astl}
\bibliography{references}

\appendix
\section{Extended Search Radius}
\label{app:catalog_check}
In the main analysis, we searched for non-thermal hard X-ray/soft $\gamma$-ray emission from Abell~3667 using INTEGRAL/ISGRI mosaics optimized for sensitivity to faint, diffuse emission. While our initial investigation extended up to 300~keV, we ultimately restricted the upper limit analysis to the 30--100~keV range. This choice is motivated by the rapidly declining ISGRI effective area above $\sim$100~keV, where instrumental background and systematic uncertainties increasingly dominate over source photons. 

To further assess the robustness of our non-detection and to exclude potential contamination from nearby hard X-ray sources, we extended the spatial search radius from the nominal $3^\circ$ region used in the main analysis to a larger $5^\circ$ radius centered on Abell~3667. All publicly available ISGRI Science Windows within this $5^\circ$ field were included, as listed in Table~\ref{tab:scw_all}. This expanded selection includes 16 Science Windows with a total good ISGRI exposure time of 22,010~s, compared to the 2,817~s used in the main analysis. The nearly eight-fold increase in exposure provides a substantially deeper dataset to verify our non-detection and search for any previously unidentified sources in the wider field.

Significance maps were generated across the full 30--100~keV energy range using four logarithmic bins, following the same procedure described in Section~\ref{sec:methods}. No previously uncataloged sources were detected with a significance exceeding $5\sigma$ in any energy band. This confirms that the non-detection reported in the main text is not an artifact of the initial spatial selection and that no transient or persistent sources emerge when the field of view is expanded. 

Within a broader $\sim25^\circ$ region centered on the cluster, we identify 26 cataloged hard X-ray sources from the INTEGRAL/ISGRI catalog. These sources are listed in Table~\ref{tab:nearby_sources}, ordered by increasing angular separation from the cluster center. To quantitatively verify the absence of contamination at the cluster position, we extracted pixel-level significance values directly from the ISGRI mosaic significance maps at the locations of Abell~3667 and the two nearest cataloged sources. These values were measured independently in each of the four logarithmic energy bands spanning 30--100~keV. The results are summarized in Table~\ref{tab:extended_significance_values}. 

In all bands, the significance at the cluster position remains well below the standard ISGRI detection threshold of $5\sigma$. The maximum significance reached is only $0.44\sigma$ in the 41--54.5~keV band. The same holds for the two nearest cataloged sources: SWIFT~J2012.0$-$5648, located just $2.8'$ from the cluster center, shows a maximum significance of $1.37\sigma$ in the 54.5--73.5~keV band, while SWIFT~J2018.4$-$5539 at $81.9'$ separation reaches $2.19\sigma$ in the same band. Neither source shows significant emission during the epochs covered by our analysis, confirming that contamination from nearby sources does not affect our cluster-centered analysis. The comprehensive nature of this extended-radius search, with its significantly increased exposure time and wider field coverage, strongly validates the robustness of our non-detection of soft $\gamma$-ray emission from Abell~3667.

\begin{table*}[ht]
\centering
\caption{Science Windows within $5^\circ$ of Abell~3667 (RA~=~$303.14^\circ$, Dec~=~$-56.84^\circ$, J2000). Angular separations are computed between the ScW pointing center and the cluster position.}
\label{tab:scw_all}
\begin{tabular}{lccc}
\hline\hline
\textbf{ScW ID} & \textbf{Pointing Center (RA, Dec)} & \textbf{good\_isgri (s)} & \textbf{Separation (arcmin)} \\
\hline
091100160010 & $(302.886^\circ,\,-54.290^\circ)$ & 506 & 153.3  \\
186500380010 & $(306.982^\circ,\,-58.417^\circ)$ & 844  & 155.5 \\
091100170010 & $(302.868^\circ,\,-54.246^\circ)$ & 688 & 155.9 \\
186500370010 & $(307.235^\circ,\,-58.605^\circ)$ & 779 & 168.6  \\
275800320010 & $(297.576^\circ,\,-57.730^\circ)$ & 635 & 188.1  \\
275800310010 & $(297.715^\circ,\,-58.061^\circ)$ & 475 & 189.8  \\
102300500010 & $(297.158^\circ,\,-56.429^\circ)$ & 1950 & 198.9  \\
102300490010 & $(297.140^\circ,\,-56.421^\circ)$ & 466 & 199.5  \\
102500860010 & $(297.286^\circ,\,-55.649^\circ)$ & 1780 & 207.7  \\
114400660010 & $(296.917^\circ,\,-58.039^\circ)$ & 1905 & 213.3  \\
108200430010 & $(296.962^\circ,\,-55.633^\circ)$ & 1896 & 218.3  \\
114400020010 & $(296.882^\circ,\,-58.308^\circ)$ & 3647 & 219.7  \\
114201040010 & $(296.774^\circ,\,-59.128^\circ)$ & 774 & 244.5 \\
114300090010 & $(296.855^\circ,\,-59.205^\circ)$ & 1953 & 244.8  \\
114300100010 & $(296.769^\circ,\,-59.159^\circ)$ & 1897 & 245.6  \\
108301060010 & $(296.343^\circ,\,-60.193^\circ)$ & 1815 & 292.7  \\
\hline\hline
\end{tabular}
\end{table*}

\begin{table*}[ht]
\centering
\caption{Cataloged hard X-ray sources within $25^\circ$ of Abell~3667. Sources are listed in order of increasing angular separation from the cluster center (RA~=~$303.14^\circ$, Dec~=~$-56.84^\circ$, J2000).}
\label{tab:nearby_sources}
\begin{tabular}{lccc}
\hline\hline
\textbf{Source Name} & \textbf{RA (deg)} & \textbf{Dec (deg)} & \textbf{Separation (arcmin)} \\
\hline
SWIFT~J2012.0$-$5648 & 303.06 & $-56.82$ & 2.8 \\
SWIFT~J2018.4$-$5539 & 304.43 & $-55.68$ & 81.9 \\
ESO~141$-$55         & 290.31 & $-58.67$ & 424.4 \\
IGR~J19387$-$6502    & 294.74 & $-65.04$ & 548.6 \\
IGR~J19386$-$4653    & 294.65 & $-46.89$ & 673.8 \\
ESO~140$-$43         & 281.22 & $-62.37$ & 738.4 \\
SWIFT~J1839.1$-$5717 & 279.76 & $-57.25$ & 759.6 \\
IGR~J21024$-$4608    & 315.63 & $-46.11$ & 792.5 \\
ESO~103$-$35         & 279.58 & $-65.43$ & 846.5 \\
IGR~J18244$-$5622    & 276.08 & $-56.37$ & 888.2 \\
IGR~J20526$-$4320    & 313.16 & $-43.35$ & 894.7 \\
QSO~B1933$-$400      & 294.32 & $-39.97$ & 1069.5 \\
IGR~J17520$-$6018    & 267.98 & $-60.33$ & 1105.3 \\
IGR~J17431$-$5945    & 265.78 & $-59.77$ & 1174.7 \\
IGR~J17379$-$5957    & 264.40 & $-59.96$ & 1214.3 \\
IGR~J19077$-$3925    & 286.96 & $-39.39$ & 1224.3 \\
IGR~J17427$-$7319    & 265.80 & $-73.32$ & 1324.2 \\
IGR~J17217$-$6030    & 260.41 & $-60.53$ & 1326.2 \\
ESO~399$-$20         & 301.74 & $-34.55$ & 1338.7 \\
NGC~6300             & 259.25 & $-62.82$ & 1342.0 \\
2E~1849.2$-$7832     & 284.28 & $-78.47$ & 1351.6 \\
ESO~25$-$G002        & 283.84 & $-78.86$ & 1374.5 \\
IGR~J17173$-$5855    & 259.33 & $-58.93$ & 1377.9 \\
IGR~J17062$-$6143    & 256.57 & $-61.71$ & 1424.4 \\
IGR~J17008$-$6425    & 255.20 & $-64.43$ & 1441.9 \\
IGR~J17063$-$7338    & 256.49 & $-73.61$ & 1477.2 \\
\hline
\hline
\end{tabular}
\end{table*}

\begin{table}[ht]
\centering
\caption{Significance values at source positions in the extended $5^\circ$ field ISGRI significance maps. Values are extracted at the positions of Abell~3667 and the two nearest catalog sources across the four analyzed energy bands.}
\label{tab:extended_significance_values}
\begin{tabular}{lcccc}
\hline\hline
Source & Band 1 & Band 2 & Band 3 & Band 4 \\
       & (30--41 keV) & (41--54.5 keV) & (54.5--73.5 keV) & (73.5--99.5 keV) \\
\hline
Abell 3667                     & 0.15 & 0.44 & 0.09 & 0.11 \\
SWIFT J2012.0-5648             & 0.00 & 0.46 & 1.37 & 0.00 \\
SWIFT J2018.4-5539             & 0.00 & 1.42 & 2.19 & 0.21 \\
\hline
\end{tabular}
\end{table}

\end{document}